\documentclass[aps,prb,twocolumn,epsfig,twoside,a4paper]{revtex4}
\usepackage{graphicx} 
\begin{document}

\title{Anomalous dynamics of confined water at low hydration.}
\author{P.~Gallo and M.~Rovere}
\address{Dipartimento di Fisica, 
Universit\`a ``Roma Tre'' \\
Istituto Nazionale per la Fisica della Materia,
Unit\`a di Ricerca Roma Tre and DEMOCRITOS National
Simulation Center \\
Via della Vasca Navale 84, 00146 Roma, Italy.}

\begin{abstract}
The mobility of water molecules confined in a silica
pore is studied by computer simulation in the low
hydration regime, where most of the molecules reside
close to the hydrophilic substrate. A layer analysis 
of the single particle dynamics of these molecules shows an 
anomalous diffusion 
with a sublinear behaviour at long time. This behaviour is
strictly connected to the long time decay of the residence time distribution
analogously to water at contact with proteins. 
\end{abstract}

\maketitle

\section{ Introduction}

Water plays a major role in many
different biological, chemical and physical
phenomena~\cite{debenedetti,robinson,grenoble}.
In most of these cases a large fraction of water is 
at contact with different substrates and its 
motion is restricted in small spaces. 

It is expected that both the geometrical confinement and the interaction
with the substrate perturb the structural and dynamical 
properties of water and some general trends
have been found in experiments and computer 
simulation~\cite{noi-grenoble03}.

In many phenomena, like those connected
with biological matter, it is of great relevance 
the behaviour of the shells of water close to substrates.
In particular the slow dynamics of water close to the 
surfaces of proteins might play a fundamental role in
the protein functionality and 
evidences have been found
of a glasslike behaviour of water at contact with plastocyanin 
investigated in a wide temperature range by 
computer simulation~\cite{bizzarri1}. 

Water confined in Vycor glass
is a good prototype system for studying the effect of
an hydrophilic substrate
since the surface of Vycor pores is well
characterized~\cite{mar1}.   
In particular
experimental studies with quasi-elastic neutron scattering
and neutron resonance spin-echo~\cite{zanotti,zanotti1} 
on confined water indicated
a slowing down of the dynamics with respect to the 
bulk and a study focused on the low hydration regime
has evidenced upon supercooling the existence
of a low frequency scattering excess typical of strong glass
former~\cite{venturini}.

In computer simulations  
of water confined in a pore of Vycor glass we found
that strong layering effects are present where the
molecules close to Vycor glass show very slow dynamics even at ambient 
temperature~\cite{noi-prl,noi-jcp,noi-jpcm}. 
In a more focused study on the low hydration regime~\cite{lowhyd,eps2002}
we recently performed also a preliminary study of
the residence time (RT) of the water molecules. We found that 
the RT is strongly dependent on the distance from the substrate
and its distribution shows an anomalous non-brownian behaviour
when the contribution of the molecules close to the substrate
alone is considered.

The aim of this paper is to show the connection between
the anomalous behaviour of the RT and the long time limit
of the mean square displacement (MSD) in the low hydration regime. 
The paper is structured as follows: in the next section
we briefly describe the system, give details of the
computer simulation and discuss some structural properties
of confined water~\cite{lowhyd} useful
for the characterization of the RT behaviour. In the third section 
by considering the residence time distribution and
the molecular diffusion we show the onset of
an anomalous behaviour connected to the presence of the solid disordered
substrate. The last section is devoted to conclusions.

\section{ Structural properties of confined water}

The molecular dynamics (MD) calculations have been performed in a cell 
of silica glass previously obtained
by the usual quenching procedure~\cite{jmol1}. 
\begin{figure}
\begin{center}
\includegraphics[clip,width=8cm]{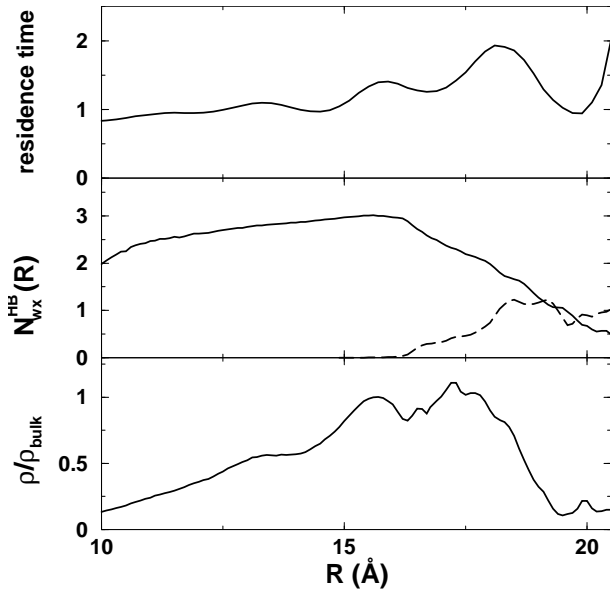}
\caption{For $N_W=1500$ water molecules at $T=300$~K
along the cylindrical radius $R$ are shown:
at the bottom density profiles along
the cylindrical radius $R$ at $T=300$~K
normalized to the density of bulk water at ambient
conditions; in the middle number of
hydrogen bonds (HB) per molecule,
the bold line represents the water-water HB,
the dotted line the water-Vycor HB; at the top  
residence time 
along the cylindrical radius $R$. 
}
\protect\label{fig:1a}
\end{center}
\end{figure} 

\begin{figure}
\begin{center}
\includegraphics[clip,width=8cm]{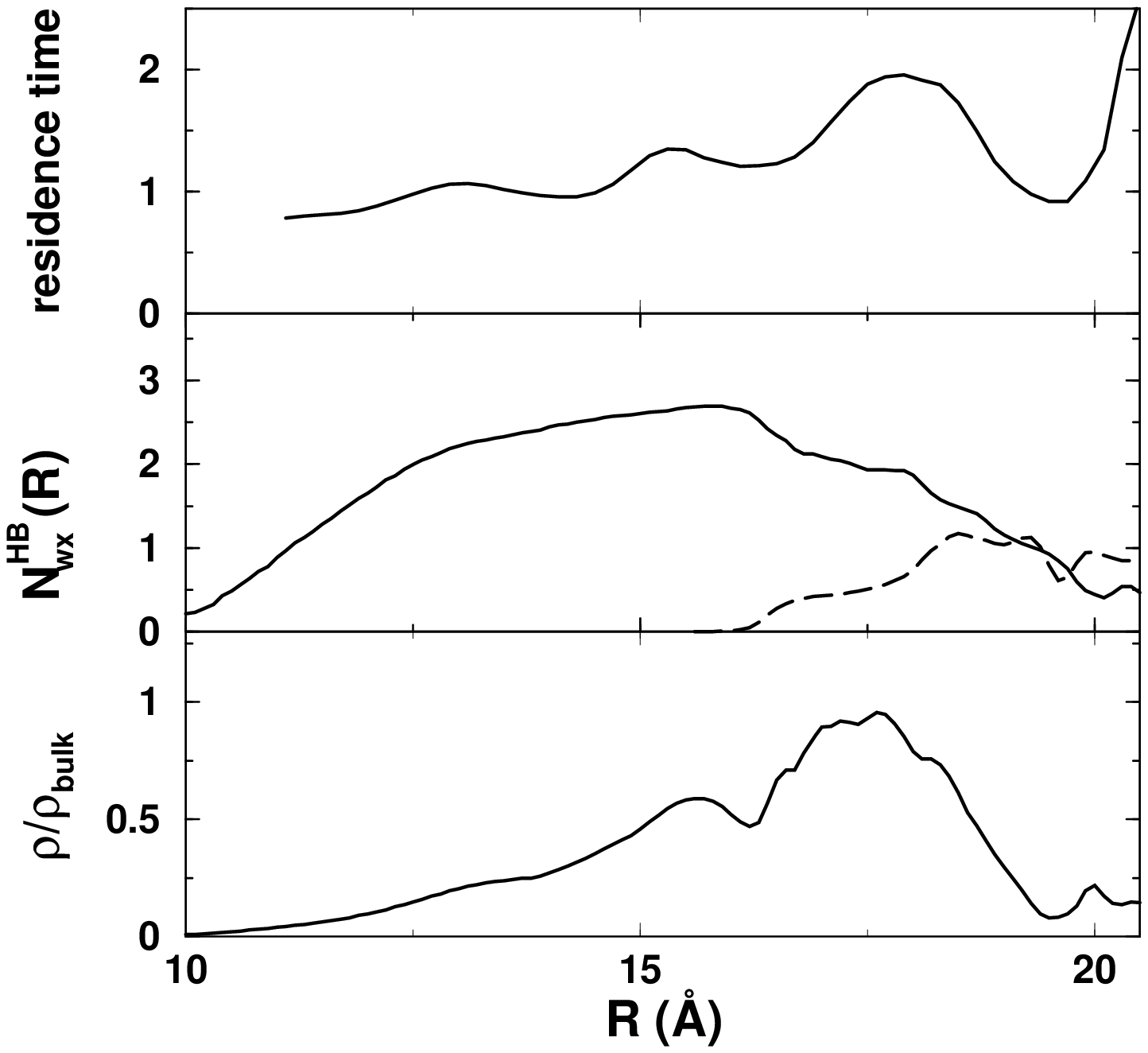}
\caption{For $N_W=1000$ water molecules at $T=300$~K
along the cylindrical radius $R$ are shown:
at the bottom density profiles along
the cylindrical radius $R$ at $T=300$~K
normalized to the density of bulk water at ambient
conditions; in the middle number of
hydrogen bonds (HB) per molecule,
the bold line represents the water-water HB,
the dotted line the water-Vycor HB; at the top  
residence time 
along the cylindrical radius $R$. 
}
\protect\label{fig:1b}
\end{center}
\end{figure} 

\begin{figure}
\begin{center}
\includegraphics[clip,width=8cm]{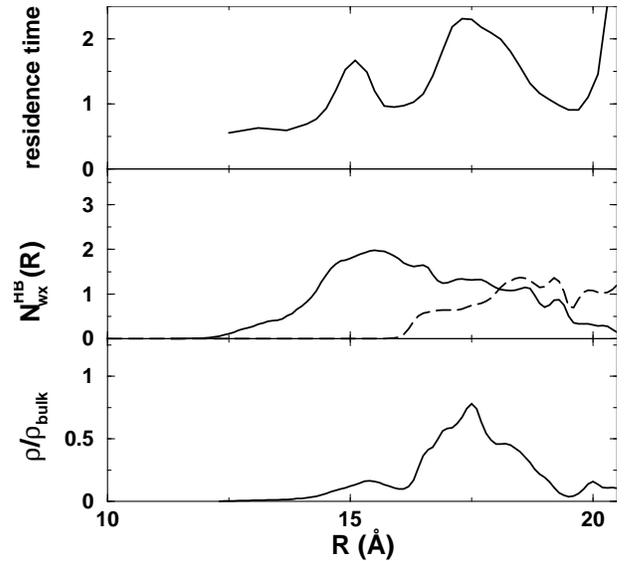}
\caption{For $N_W=500$ water molecules at $T=300$~K
along the cylindrical radius $R$ are shown:
at the bottom density profiles along
the cylindrical radius $R$ at $T=300$~K
normalized to the density of bulk water at ambient
conditions; in the middle number of
hydrogen bonds (HB) per molecule,
the bold line represents the water-water HB,
the dotted line the water-Vycor HB; at the top  
residence time 
along the cylindrical radius $R$. 
}
\protect\label{fig:1c}
\end{center}
\end{figure}

Inside the glass cubic cell of $71$~\AA\ a 
cylindrical cavity of $40$~\AA\ diameter
has been carved. The surface of the cavity has been treated
to reproduce the average properties of the surface
of the pores of Vycor. Along this line
hydrogen ions (acidic hydrogens) have been
added to oxygens non saturated by silicons in order to 
mimic the procedure followed by experimentalists in the
preparation of the Vycor sample before hydration.  
The surface of silica glass can be considered
as a prototype model for a disordered hydrophilic substrate. 
Being primarily interested in the dynamics
of water the substrate is kept rigid. 
The water inserted in the cavity is simulated by using
the SPC/E potential, where each molecule is
represented by three charged sites. These sites interact also with the
silicon and oxygen atoms of the substrate by means of an hydrophilic
potential described in previous work~\cite{noi-jcp,jmol1}.
The molecular dynamics (MD) is performed at different hydrations
by varying the number of water molecules contained in the pore. 
For each hydration
the system is equilibrated at different 
temperatures. The quantities of interest presented in the
following are averaged over runs that extend up to $1.2-1.3$~ns.
We note that we have extended the simulation length at all 
temperatures with
respect to ref.~\cite{lowhyd} to improve
statistics especially on the computation of the RT. 

The number of water molecules considered 
in this work are $N_W=500$,
$N_W=1000$, $N_W=1500$, which correspond to hydration levels
of the pore of $19 \%$, $38 \%$ and $56 \%$ respectively,  
since the density corresponding to
the full hydration in the experiments $\rho=0.878$~g/cm$^2$ 
is obtained in our geometry for $N_W=2600$.

The effects of the hydrophilic interaction of the substrate on the
water molecules are shown in the bottom panels of
Fig.~\ref{fig:1a},~\ref{fig:1b} and~ \ref{fig:1c} for the different
hydrations. 
The radial density profiles normalized to the density of
bulk water at ambient conditions show the formation of two
layers of molecules close to the substrate. 
The positions of the peaks of
the double layer structure do not change with hydrations,
the first layer is located at around $R=17.5$~\AA\ and the second layer is
at $R \approx 15.5$~\AA\ with a minimum in the density at
$R \approx 16$~\AA. The heights of the peaks increase with increasing 
hydrations and
for $N_W = 1500$ the normalized density profile for confined water in the
layers reaches values  higher than one at ambient temperature.     

In the middle panels of Fig.~\ref{fig:1a},~\ref{fig:1b} and~\ref{fig:1c}
it is shown that
the intermolecular hydrogen bond (HB) profiles increase and
reach a maximum value at the position of the minimum of density
($R \approx 16$~\AA) where they start to go down in correspondence with
the increase of the HB of water molecules with the atoms of the
Vycor surface. The layering effect shown in the bottom panels 
is due to the formation of the Vycor-water HB. 
We found that the temperature has little effects
on the density and HB profiles at all the hydration levels
and for this reason we show here only the results corresponding to ambient 
temperature.


\section{Residence time and anomalous diffusion}

At the top of Fig.~\ref{fig:1a},~\ref{fig:1b} and~\ref{fig:1c}
the residence times (RT) of the water molecules
at ambient temperature are reported along the pore radius~\cite{lowhyd}.
The large oscillations of the RT appear modulated by
the structure of the density profiles,
reported in the bottom panels of the same figures.
Apart for the molecules attached to the surface
water resides for the longer time inside the shells,
where the density profile reaches the highest values.
The minima of the RT are located close to the minima
of the density profiles.

\begin{figure}
\begin{center}
\includegraphics[clip,width=8.5cm]{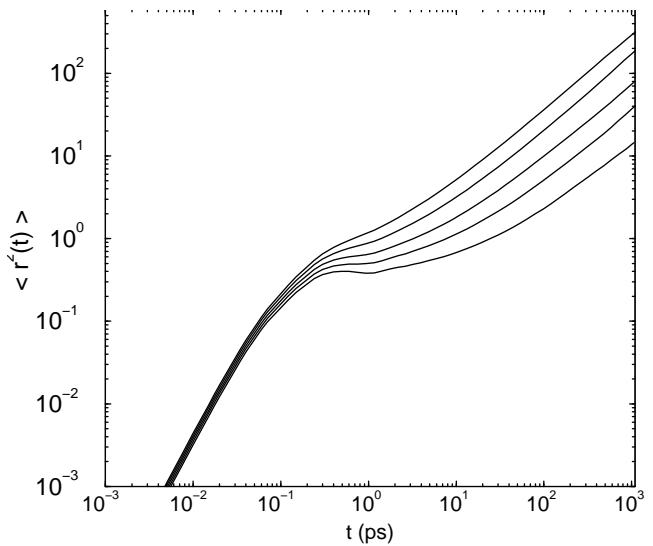}
\end{center}
\caption{ MSD for $N_W=1000$ at decreasing temperature:
$T=300,280,260,240,220$~K from above.
}
\protect\label{fig:2}
\end{figure}

In Fig~\ref{fig:2} are reported the calculations of the mean
square displacement (MSD) at $N_W=1000$ for decreasing temperatures.
It is clear that after the ballistic regime at short time, at around $0.1$~ps
there is the onset of a cage effect characterized by the presence
of a plateau which increases by lowering the temperature.
The plateau is determined by the transient caging of the 
nearest neighbours.
At longer times the MSD does not appear to reach the usual Brownian diffusion
since the behaviour is sub-linear. 
Analogous results are found for the other hydration levels investigated.
At this point further analysis is needed in order to clarify whether
this subdiffusive behaviour is just a transient leading to
normal brownian diffusion for longer times, although unreachable
with normal computers, or it can be framed
in the context of anomalous subdiffusive phenomena.

Anomalous diffusion is generally speaking defined trough the 
time dependence of the MSD, the second moment of the spatial coordinates
of the diffusive particles, which generally has a long time dependence
of the form 
\begin{equation}
<r^2(t)>=a t^\alpha
\end{equation}
Anomalous diffusion corresponds to $\alpha \ne 1$. In particular
$\alpha > 1$ is termed superdiffusion and $\alpha < 1$ subdiffusion.
From a theoretical point of view the origin of anomalous diffusion
can be traced back to the analytic form of the distribution of the
waiting times $\psi(t)$. 
Under the assumption that the amplitude of the random jumps
is constant and finite anomalous diffusion has been shown 
to be generated by an inverse power law distribution
for large times~\cite{mannella,montroll} 
\begin{equation}
\psi(t)=A t^{-\mu} \label{eq:1}
\end{equation}
In the case of ordinary brownian motion the long time limit
of the distribution
would decay with an exponential law.

In our case the sublinear
diffusion observed can be connected to the 
processes which take place close to the substrate and to the
interaction of the water molecules with the disordered surface~\cite{bychuk}.
In this respect since
oscillations of the RT appear so closely connected to the double
layer structure it is of interest
to look at the residence time distribution
of the water molecules close to the substrate.

\begin{figure}
\begin{center}
\includegraphics[clip,width=8.5cm]{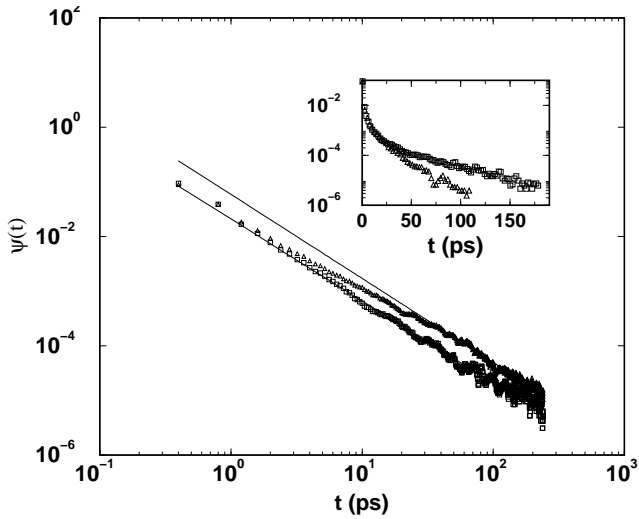}
\end{center}
\caption{Log-log plot of the residence time distribution (RTD)
of the water molecules
in the layer $14<R<20$~\AA\ for $N_W=1000$ at temperatures
$T=300$~K (open triangles) 
and $T=240$~K (open squares). The fits (bold lines) 
are done with a power law $At^{-\mu}$.
$A=0.059$, $\mu=1.54 \pm 0.05$ at $T=300$~K. 
$A=0.021$, $\mu=1.50 \pm 0.05$ at $T=240$~K.
In the inset are shown in a linear-log scale 
the RTD for the molecules in the layer $0<R<14$~\AA.
}
\protect\label{fig:4}
\end{figure}

\begin{figure}
\begin{center}
\includegraphics[clip,width=8.5cm]{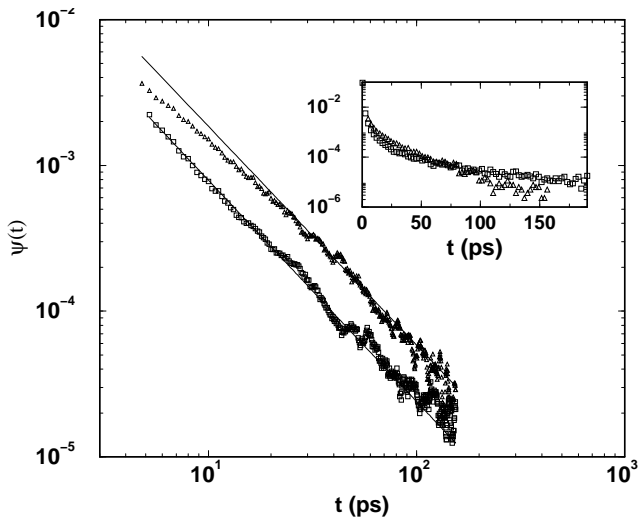}
\end{center}
\caption{Log-log plot of the residence time distribution (RTD)
of the water molecules
in the layer $14<R<20$~\AA\ for $N_W=1500$ at temperatures
$T=300$~K (open triangles)
and $T=240$~K (open square). The fits (bold lines) are
done with a power law $At^{-\mu}$.
$A=0.059$, $\mu=1.50 \pm 0.05$ at $T=300$~K. 
$A=0.026$, $\mu=1.52 \pm 0.05$ at $T=240$~K.
In the inset are shown in a linear-log scale the
RTD for the molecules in the layer $0<R<14$~\AA.
}
\protect\label{fig:5}
\end{figure}

In Fig.~\ref{fig:4} and \ref{fig:5} 
we report the residence time distributions (RTD) $\psi(t)$
at the highest $T=300$~K and the lowest $T=240$~K 
temperatures investigated for $N_W=1000$ and $N_W=1500$ . 
The RTD, calculated for the molecules in the layer $14<R<20$~\AA, shows
the predicted power law behaviour of Eq.~\ref{eq:1}
while for the rest 
of the molecules we get an exponential decay, 
as shown in the inset of the figures.
The power law behaviour related to the temporal disorder of the
distribution of the residence times of the molecules
has been observed in computer simulation and experiments on
water at contact with proteins~\cite{bizzarri1}. They are
specifically related to the interaction of the solvent with the protein
sites. 

The power law decay
of the RTD of the molecules in the $6$~\AA\ layer from
the surface is determined by values of the
exponents which are similar to the ones obtained for the RTD of 
water in few \AA\ shells close to protein hydration sites~\cite{bizzarri1}.
In particular for the case $N_W=1000$ we have 
$\mu=1.54 \pm 0.05$ at $T=300$~K and $\mu=1.50 \pm 0.05$ at $T=240$~K,
while for $N_W=1500$ the fits yield $\mu=1.50 \pm 0.05$ at $T=300$~K 
and $\mu=1.52 \pm 0.05$ at $T=240$~K. We note that the 
present result for
$\mu$ at $N_W=1500$ and ambient temperature is slightly different
from the preliminary one reported in ref.~\cite{lowhyd}, where
the statistics was poorer.

The sublinear behaviour of the MSD is connected to
the power law decay of the RTD by the asymptotic temporal
dependence
\begin{equation}
<r^2(t)> \propto t^{\mu-1} \label{eq:2}
\end{equation}

\begin{figure}
\begin{center}
\includegraphics[clip,width=8.5cm]{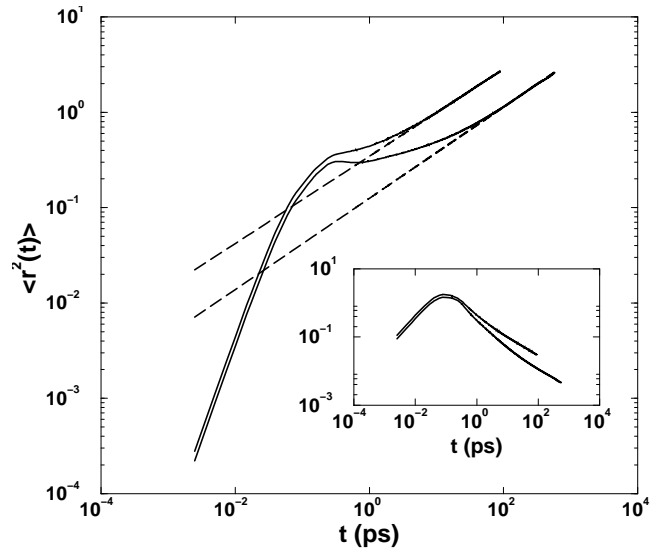}
\end{center}
\caption{MSD of water
molecules in the layer $14<R<20$~\AA\ for $N_W=1000$ at 
temperatures $T=300$~K and $T=240$~K from above. 
The long dashed lines are 
the fits to a sublinear behaviour $<r^2> \propto t^\alpha$ with
$\alpha=0.46 \pm 0.05$ at $T=300$~K and $\alpha=0.48 \pm 0.05$
at $T=240$~K.
In the inset are reported the functions $<r^2>/t$. 
}
\protect\label{fig:6}
\end{figure}

\begin{figure}
\begin{center}
\includegraphics[clip,width=8.5cm]{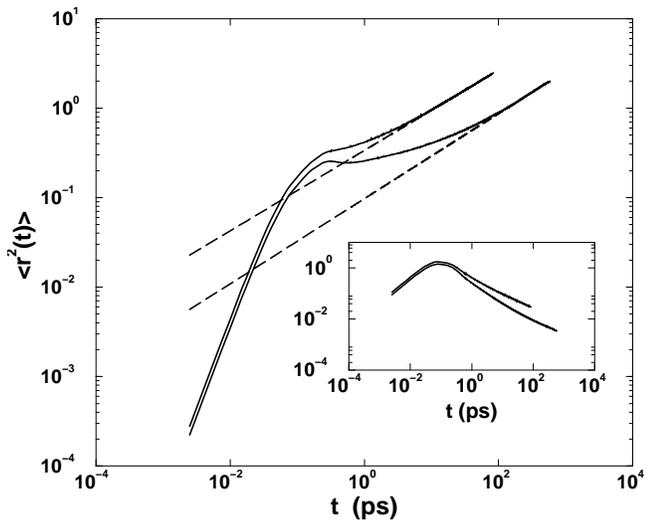}
\end{center}
\caption{ MSD of water
molecules in the layer $14<R<20$~\AA\ 
for $N_W=1500$ at temperatures $T=300$~K and $T=240$~K from above. 
The long dashed lines are 
the fits to a sublinear behaviour $<r^2> \propto t^\alpha$ with
$\alpha=0.45 \pm 0.05$ at $T=300$~K and $\alpha=0.48 \pm 0.05$
at $T=240$~K.
In the inset are reported the functions $<r^2>/t$. 
}
\protect\label{fig:7}
\end{figure}

From the fit of the long time behaviour of the MSD for $N_W=1000$
and $N_W=1500$, reported in Fig.~\ref{fig:6} and Fig.~\ref{fig:7}, we get
the values of the exponent $\mu$ which are 
consistent with the exponents obtained from the power law
behaviour of the RTD seen in Fig.~\ref{fig:4} and Fig.~\ref{fig:5}. 

\begin{figure}
\begin{center}
\includegraphics[clip,width=8.5cm]{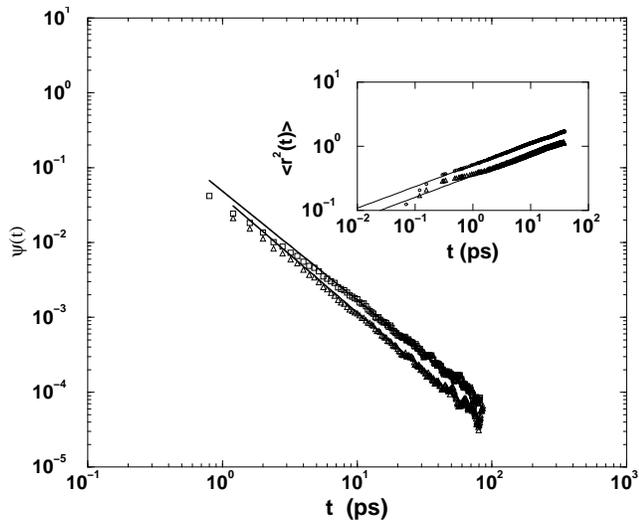}
\end{center}
\caption{ Log-log plot of the residence time distribution (RTD)
of the water molecules
in the layer $14<R<20$~\AA\ for $N_W=500$ at temperatures
$T=300$~K (open squares) 
and $T=240$~K (open triangles). 
The fit (bold lines) 
is done with a power law $At^{-\mu}$, with
$A=0.048$, $\mu=1.45 \pm 0.05$ for $T=300$~K and
$A=0.041$, $\mu=1.55 \pm 0.05$ for $T=240$~K.
In the inset MSD of water
molecules in the layer $14<R<20$~\AA\ 
for $N_W=500$ at temperatures $T=300$~K and $T=240$~K. 
The bold line is 
the fit to a sublinear behaviour $<r^2> \propto t^\alpha$ with
$\alpha=0.34 \pm 0.05$ for both $T=300$~K and 
$T=240$~K.
}
\protect\label{fig:8}
\end{figure}

In the lower hydration case $N_W=500$, Fig.~\ref{fig:8},
the RTD decays for long time with an exponent similar
to the previous cases, $\mu = 1.45 \pm 0.05$
and $\mu = 1.55 \pm 0.05$ for $T=300$~K and $T=240$~K respectively, but 
the MSD show, as seen  in the inset of Fig.~\ref{fig:8},
a long time behaviour not in agreement 
with the one predicted by Eq.~\ref{eq:2}.

The asymptotic behaviour of the MSD shows a further 
slowing down with respect to the
higher hydrations. This behaviour is likely connected 
to the fact that the
water molecules are arranged in clusters close to the
substrate. 

\section{Conclusions}

The dynamical properties of confined water are
expected to be changed by the interaction with the 
substrate. We performed
computer simulation of water molecules
confined in a silica pore in 
low hydration regimes, where the larger amount
of water resides in the shells closer to the hydrophilic surface.
We found drastic changes of the diffusion of
the water molecules.
From a layer analysis 
for the investigated hydrations
we find that the diffusive regime at long time
of the molecules close to the substrate 
is characterized by a sublinear trend.

Anomalous diffusion phenomena are
related to a temporal disorder typical of particles which 
diffuse close to and interact with 
a disordered surface. Different 
interaction processes between the water molecules and
the sites of the substrate modulate the residence time
of the molecules~\cite{bizzarri1}.
A dispersive transport regime related to 
temporal disorder shows up in the power law decay of
the residence time distribution with an exponent which
determines also the long time tail of the mean square 
displacement. In our system
the exponent
of the long time behaviour of the mean square displacement
is related to the long time decay of the
residence time distribution of the molecules in the
same layer for the cases $N_W=1500$ and $N_W=1000$
as theoretically predicted.
For the lowest hydration case ($N_W=500$) the 
mobility of the molecules is more strongly modulated
by the substrate with respect to the higher hydrations.
The formation of clusters
of molecules close to the solid surface does not appear to modify 
the long time decay of the RTD 
but it induces a further slowing
down of the dynamics with a violation of the expected 
behaviour of the MSD. 
\section {Acknowledgments}
We thank J. Baschnagel for useful and stimulating discussions.

\end{document}